\documentclass[twocolumn,twoside]{ncnsd}
\newcommand{\beq}{\begin{equation}}
\newcommand{\eeq}{\end{equation}}
\usepackage{latexsym}
\begin{document}
%\vspace{1cm}
%\begin{center}
%\title
%\maketitle 
\title{A phase diagram for spatio-temporal intermittency in the sine circle map
lattice }
\author{Zahera Jabeen$^{1}$ and Neelima Gupte$^{2}$
\thanks{Both authors work at the Department of Physics, Indian Institute of Technology, Madras. We thank CSIR for partial financial support. e-mail: 1. zahera@physics.iitm.ac.in, 2. gupte@physics.iitm.ac.in}}
\markboth{NATIONAL CONFERENCE ON NONLINEAR SYSTEMS \& DYNAMICS}{INDIAN INSTITUTE OF TECHNOLOGY, KHARAGPUR 721302, DECEMBER 28-30, 2003}
\maketitle

\begin{abstract}

We study the phase
diagram of the sine circle map lattice  with random initial conditions 
and identify the various types of dynamical
behaviour which appear here. We focus on the regions which
show spatio-temporal intermittency and characterise the accompanying
scaling behaviour. Directed percolation exponents are seen at some
points in the parameter space in the neighbourhood of bifurcation
boundaries. We discuss this behaviour as well as other types of
behaviour seen in the parameter space in the context of the phase
diagram.
\end{abstract}
\section{Introduction}

\PARstart{T}{he} existence of spatio-temporal intermittency wherein laminar regions,
which exhibit regular dynamics in space and time, co-exist and propagate
together with regions which show irregular or chaotic bursts, which can be termed turbulent behaviour, can be
seen in a wide variety of spatially extended systems. Examples
of such systems range from  
oscillators, systems which show pattern formation
\cite{kaneko},
chemical reactions \cite{chem},
 to 
turbulence \cite{turb} and fluid flows \cite{Colovas},
\cite{Ciliberto}, \cite{Mutabazi}. 
The phenonomenon of temporal
intermittency has been
studied extensively and is relatively well understood \cite{berge}.
 The nature
of intermittency in spatially extended dynamical systems, however has
not
been understood very  well. The presence of spatial as well as temporal
intermittency
has implications for understanding the physics of pattern formation and
for understanding
the ubiquitous presence of structures in chaotic systems.

Coupled map lattices, i.e. 
systems with continuous variables which evolve on discrete space-time,
are particularly simple paradigms for
systems with extended spatial dimension and  show a wide range of
interesting dynamical behaviour 
\cite{kaneko}. Spatio-temporal intermittency has been observed in a 
variety of coupled map lattices \cite{chate,bohr,grassberger,janaki}.
Phase transitions to spatio-temporal intermittency are a topic of
current interest, and the identification of the universality class of
this transition has led to much discussion in the literature.
Specifically, the type of spatio-temporal intermittency where there
is no spontaneous creation of bursts, and a given laminar site can only
become turbulent if infected by turbulent neighbours, has been
conjectured to lie in the same universality class as directed
percolation, with the laminar sites being identified as the `dry' or
absorbing states, and the turbulent sites being the `wet' or percolating
states. In another type of spatio-temporal intermittency, laminar sites
can spontaneously become turbulent. These  two types of intermittency
have 
been called Type I and Type II spatio-temporal intermittency in the 
literature, respectively.

Recently, the existence of spatio-temporal intermittency has been
observed at certain parameter values for the sine circle map
lattice, a popular model for the behaviour of mode-locked
oscillators \cite{janaki}. However, the detailed phase diagram of this 
model has not yet been obtained. In this paper we study the phase
diagram of this model and identify the various types of dynamical
behaviour which appear here. We next concentrate on the regions which
show spatio-temporal intermittency and characterise the accompanying
scaling behaviour. Directed percolation exponents are seen at some
points in the parameter space in the neighbourhood of bifurcation
boundaries. We discuss this  behaviour as well as other classes of
dynamical  behaviour seen in the context of the bifurcation behaviour as well as the phase diagram
of the system.

\section{The Model}

In this paper, we consider the coupled circle map lattice
\cite{nandini} which has been used to model mode-locking behaviour.
The coupled sine circle map lattice is defined by the evolution
equations \cite{gauri}:
\beq
x_{n+1}(i)= (1-\epsilon)f(x_{n}(i)) + \frac{\epsilon}{2}[f(x_n(i+1))+f(x_n(i-1)) ]{(\rm mod \ \ 1)}
\eeq

\noindent where $t$ is the discrete time index, and $i$ is the site
 index: $i=1, \dots L $, with $L$ being the system size. The parameter
 $\epsilon$ gives the strength of the diffusive coupling between site
 $i$ and its two neighbours. The local on-site map is given by
\begin{equation}
f(x)~=~ x + \Omega -\frac{K}{2 \pi} \sin(2 \pi x)
\nonumber
\end{equation}
where the parameter $K$ is  the strength of the nonlinearity, and
$\Omega$ is the frequency of the single sine circle map in the absence
of nonlinearity. This CML has been
studied extensively and  has a rich phase
diagram with many types of attractors and strong sensitivity to
initial conditions\cite{gauri,gauri1}.
In particular, this system also
has regimes of spatio-temporal intermittency (STI) when evolved
in parallel with random initial conditions \cite{janaki}. 
For weak values of nonlinearity, i.e. for $K=1.0$ the system shows
spatio-temporal intermittency with a unique absorbing state at the
parameter values $\Omega=0.064,\epsilon=0.63775$, and $\Omega=0.068,
\epsilon=0.73277$. These mark
the transition from a laminar phase to STI. 
The laminar phase here corresponds to the synchronised fixed point
$x^{\star} = \frac{1}{2\pi}
\sin^{-1}(\frac{2\pi \Omega}{K})$, and the turbulent sites are those
which are different from $x^{\star}$.
Spatio-temporal intermittency has also been observed for 
 high values of
nonlinearity $K=3.1$, this time with infinitely many absorbing states.
An earlier study of the inhomogeneous logistic map lattice had shown
spatio-temporal intermittency in the neighbour-hood of bifurcations from
the synchronised fixed point of the system. It is therefore worthwhile
to investigate the detailed phase-diagram of the present system,
identify various types of dynamical behaviour, and correlate the
observed behaviour, especially spatio-temporally intermittent behaviour, 
with the known bifurcations that occur in the system.

\section{The phase diagram}
\begin{figure}[!t]
\centering
\includegraphics[scale=0.65]{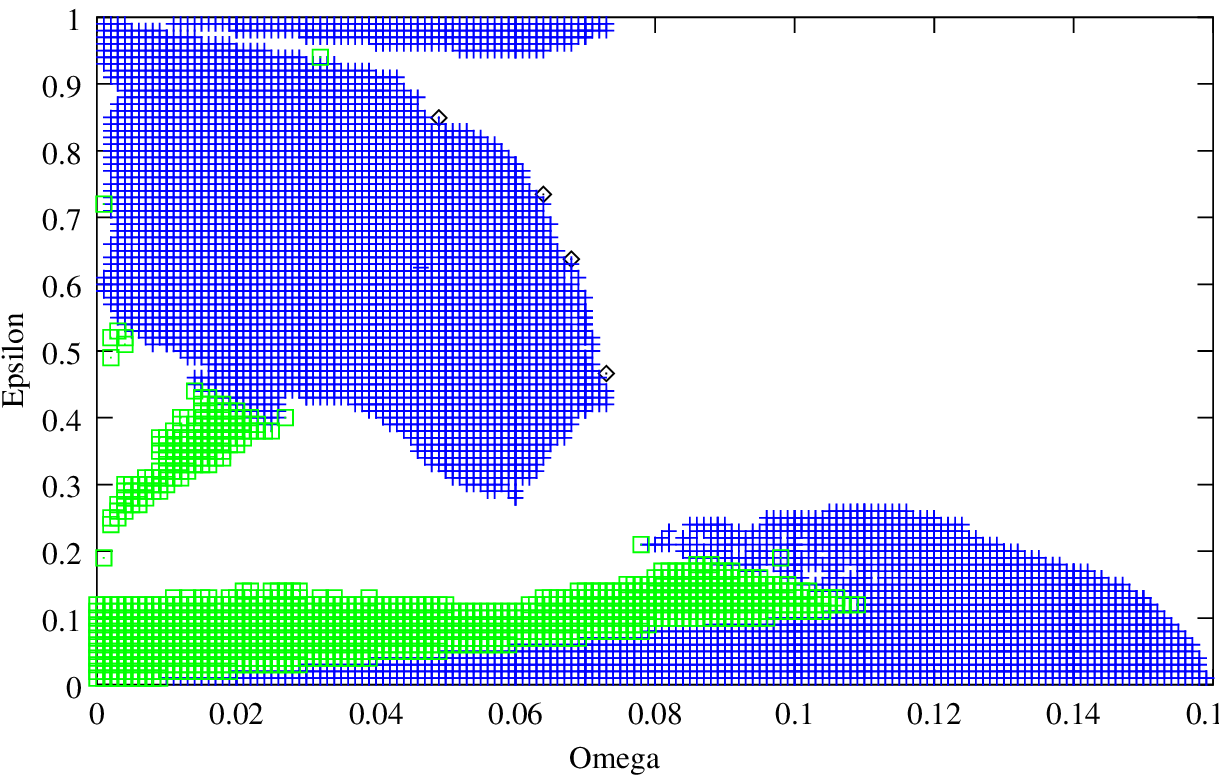}
\caption{The phase diagram in $(\epsilon,\Omega)$ space for the coupled circle map lattice after 15000 iterations for a lattice of $1000$ sites. Random initial conditions were used. The regimes of spatially synchronised and temporally fixed solutions (+) and the cluster solutions, where the number of clusters is less than 10 $(\Box)$ are shown . The points at which DP-like  exponents are obtained are also marked $(\Diamond)$.}
\end{figure}

We study the system with random initial conditions at the parameter
value $K=1.0$ as detailed phase diagrams with other classes of initial
conditions are available at this parameter value
\cite{gauri,gauri1,nandini}. The single circle map shows temporal period
1 solutions in the region $0 \le \Omega \le \frac{1}{2 \pi}$ and the
coupled sine circle map shows synchronised, period $1$ solutions over
the same range when evolved with synchronised initial conditions
\cite{nandini}. There is a similar region between $ 1- \frac{1}{2 \pi} \le
1$. We investigate this region for $0 \le \epsilon \le 1$ with random
initial conditions. The phase diagram for this region can be seen in
Fig. 1.  The following types of dynamical behaviour can be found here.

\begin{enumerate}

\item Synchronised solutions: These constitute the dominant behaviour
over the bulk of the parameter space and are very stable to perturbation.
These are indicated by plus signs in Fig. 1.

\item Cluster solutions: There are regions which correspond to clusters
where the variables on the lattice take up a finite number of values,
but at random sites in the lattice. e.g. a $2-$ cluster is a solution
where the lattice variables  take up only two values, and sites which
have the same value for the variable, belong to the same cluster.
Regions which have less than  $10$ clusters are indicated by boxes in the
phase diagram.

\item Spatio-temporally intermittent solutions can be seen at the
parameter values marked with diamonds in this region. We discuss
these solutions in detail in the next section.

\item Spatio-temporally disordered solutions can be
seen in the white regions in the parameter space.

\end{enumerate}

It is clear that the synchronised solutions change to cluster solutions
at some of the phase boundaries, and to spatio-temporally disordered
regions at others. There are some regions where a partially ordered
phase is seen in the phase diagram where there are many clusters. The
boundary between this partially ordered phase and the spatio-temporally 
disordered phase is not clear. 
While there are various interesting aspects to the behaviour of the
cluster solutions, we will concentrate on the behaviour of the
spatio-temporally  intermittent solutions  in this paper. 
 
\section{Spatio-temporal intermittency and directed percolation
exponents}

\begin{center}
\begin{figure}[!t]
\centering
\includegraphics[scale=0.7]{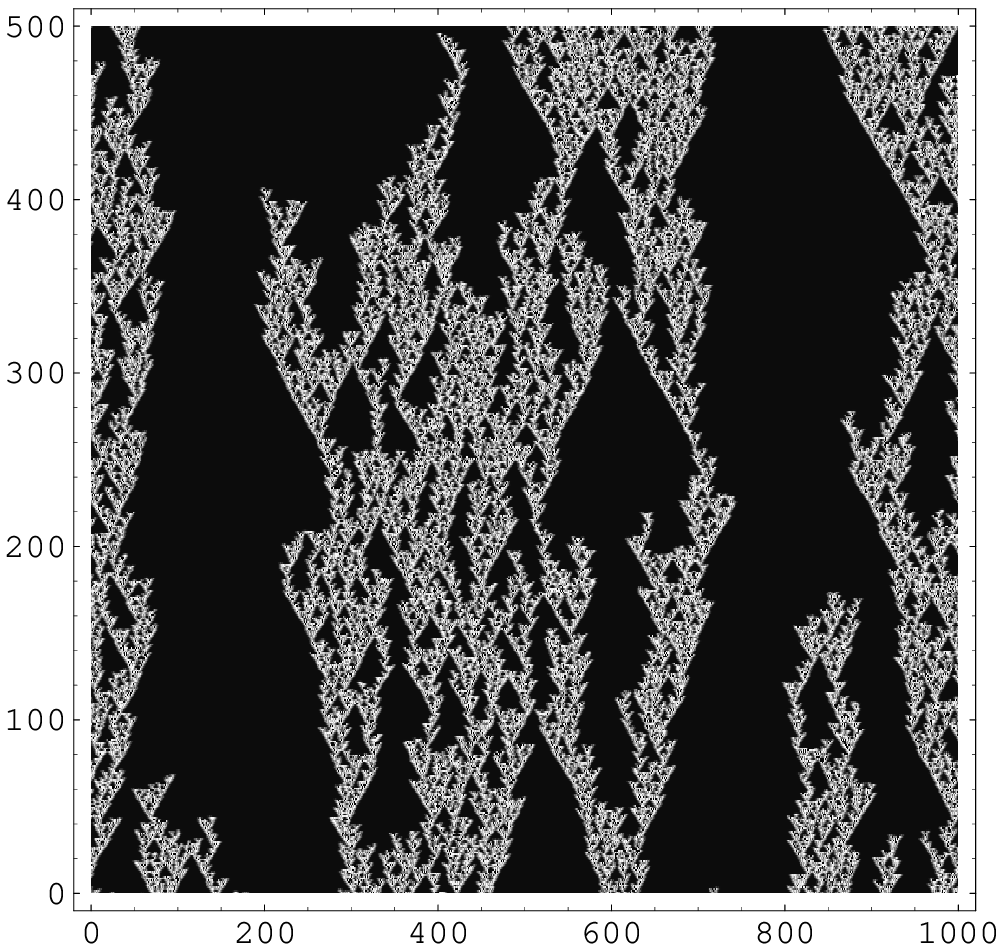}
\caption{A space-time plot for the spatio - temporal Intermittency seen in the coupled circle map lattice at $\Omega=0.049,\epsilon=0.8495$. Horizontal axis is space and time is along the vertical axis. The system size chosen is $L=1000$ and $15000$ iterations have been discarded. The black region corresponds to the laminar state and the white region corresponds to the turbulent state. Every $10^{th}$ iterate has been plotted.}
\end{figure}
\end{center}
Spatio-temporally intermittent regions can be seen at several points 
in the parameter space. The points at which spatio-temporal
intermittency of the first type are seen are marked on the phase diagram.
As mentioned earlier,  
$\Omega=0.064,\epsilon=0.63775$, and $\Omega=0.068,
\epsilon=0.73277$ are two of these points with a unique absorbing state
with the value 
$x^{\star} = \frac{1}{2\pi}
\sin^{-1}(\frac{2\pi \Omega}{K})$. The same kind of spatio-temporally
intermittent behaviour is found at two more points in the parameter
space viz. $\Omega=0.049, \epsilon=0.8494$ and $\Omega=0.073,
\epsilon=0.4664$. To verify whether the spatio-temporal intermittency 
here belongs to the universality class of directed percolation,
set of critical exponents
which describe the scaling behaviour of the quantities of physical
interest have been calculated.  The physical quantities of interest for such systems are
(a) the escape time $\tau$, which is the number of time steps elapsed
before the system reaches its laminar state and (b) the order
parameter, $m(\epsilon, L, t)$, which is the fraction of turbulent
sites in the lattice at time $t$.  From finite-size scaling arguments,
it is expected that $\tau$ depends on $L$ such that
 \[ \tau ( \Omega, \epsilon ) =\left \{
\begin{array}{ll} log~L & \mbox{laminar phase}\\ L^z &\mbox{critical
phase}\\ exp~ L^c & \mbox{turbulent phase} \end{array} \right. \]
Here, $c$ is a constant of order unity, and  the critical point is
identified as the set of parameter values at which $\tau$ shows
power-law behaviour with $z$ being the critical exponent.
\begin{figure}[!t]
\centering
\includegraphics[scale=0.6]{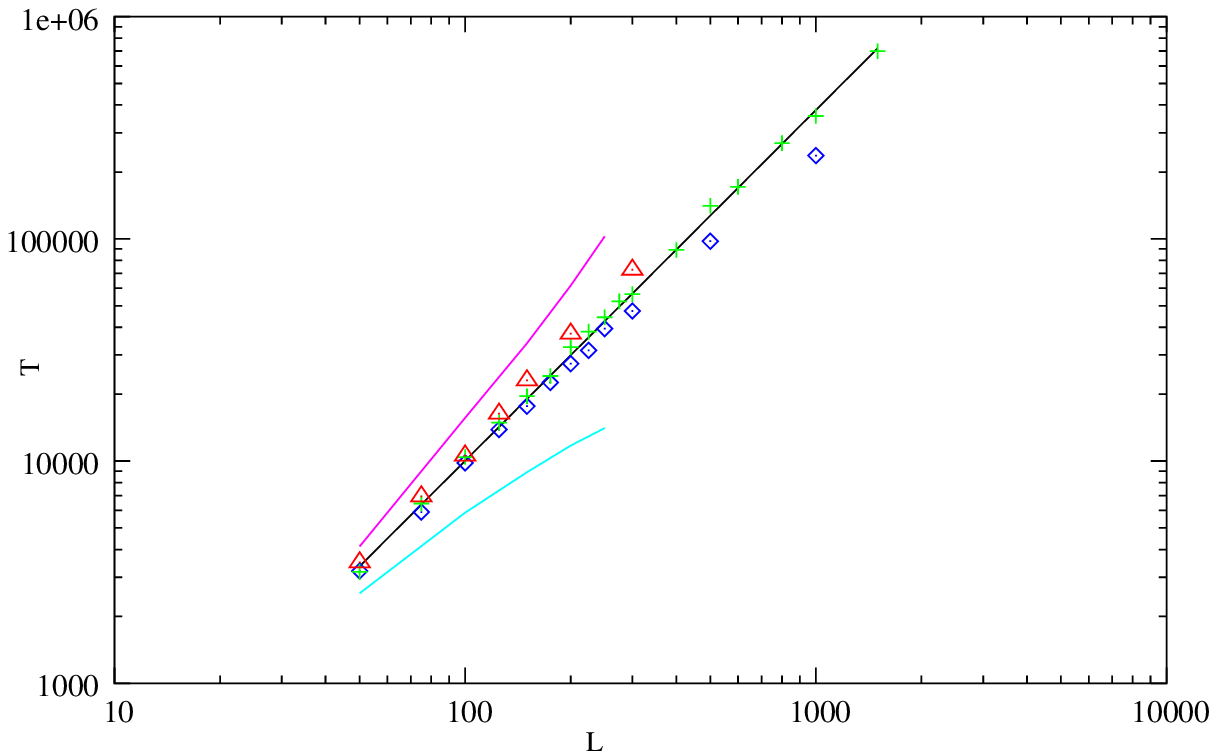}
\caption{The escape time, $\tau$ vs the system size, $L$ is plotted at $\epsilon$ values below, at and above criticality at $\Omega=0.049$ on a $\log-\log$ scale. The values of $\epsilon$ plotted are $0.845, 0.849, 0.8494, 0.85, 0.852$. Power law behaviour is seen at $\epsilon_c=0.8494$(+). The data at $\epsilon_c$ is fitted to an exponent $1.578$.} 
\end{figure}
At $\epsilon_c$ the critical value of the parameter $\epsilon$ (other
parameters being held fixed), the order parameter $m(\epsilon,L,t)$
scales as
\begin{equation}
m \sim (\epsilon-\epsilon_c)^{\beta} ,~~\epsilon \rightarrow
\epsilon_c^{+}.
\end{equation}
when the critical line is approached from above .

The order parameter is expected to obey the scaling relation
\begin{equation}
m(\epsilon_c, L,t) \approx  t^{-\beta / \nu  z}
\end{equation}

for $ t << \tau$. The exponent $\nu$ can thus be extracted once $\beta $
and $z$ are obtained from the relations above.

To extract further critical exponents, we obtain the
correlations from the pair correlation function given by:
\begin{equation}
C_j(t)=\frac{1}{L} \sum_{i=1}^L < x_i(t) x_{i+j}(t) > - < x_i(t) >^2
\end{equation}
where the brackets denote the averaging over different
initial conditions. At criticality one expects an algebraic decay of
correlation \cite{corr}:
$$C_j(t) \approx j^{1- \eta^{\prime}}$$
where $\eta^{\prime}$ is the
associated critical exponent. 

The above set of exponents constitute the static exponents of the model.
The dynamic exponents are given by the spreading exponents defined as
follows. We consider temporal evolution from initial conditions which 
correspond to an absorbing background with a localised disturbance, i.e.
a few contiguous sites which are different from an absorbing background.
The quantities of interest are, the time dependence of $N(t)$,
the number of active sites at time $t$ averaged over all runs, $P(t)$,
the survival probability, or the fraction of initial conditions which
show a non-zero number of active sites (or a propagating disturbance)
at time $t$ and the radius of gyration $R^2(t)$, which is defined as the mean squared deviation of position of the active sites from the original sites of the turbulent activity, averaged over the surviving runs alone. 
At
criticality, we have,
$$N(t) \approx t^{\eta},~~P(t) \approx t^{- \delta},~~R^2(t) \approx
t^{z_s}.$$
Also, $\delta=\beta / \nu z$.
\begin{center}
\begin{figure}[!t]
\centering
\includegraphics[scale=0.7]{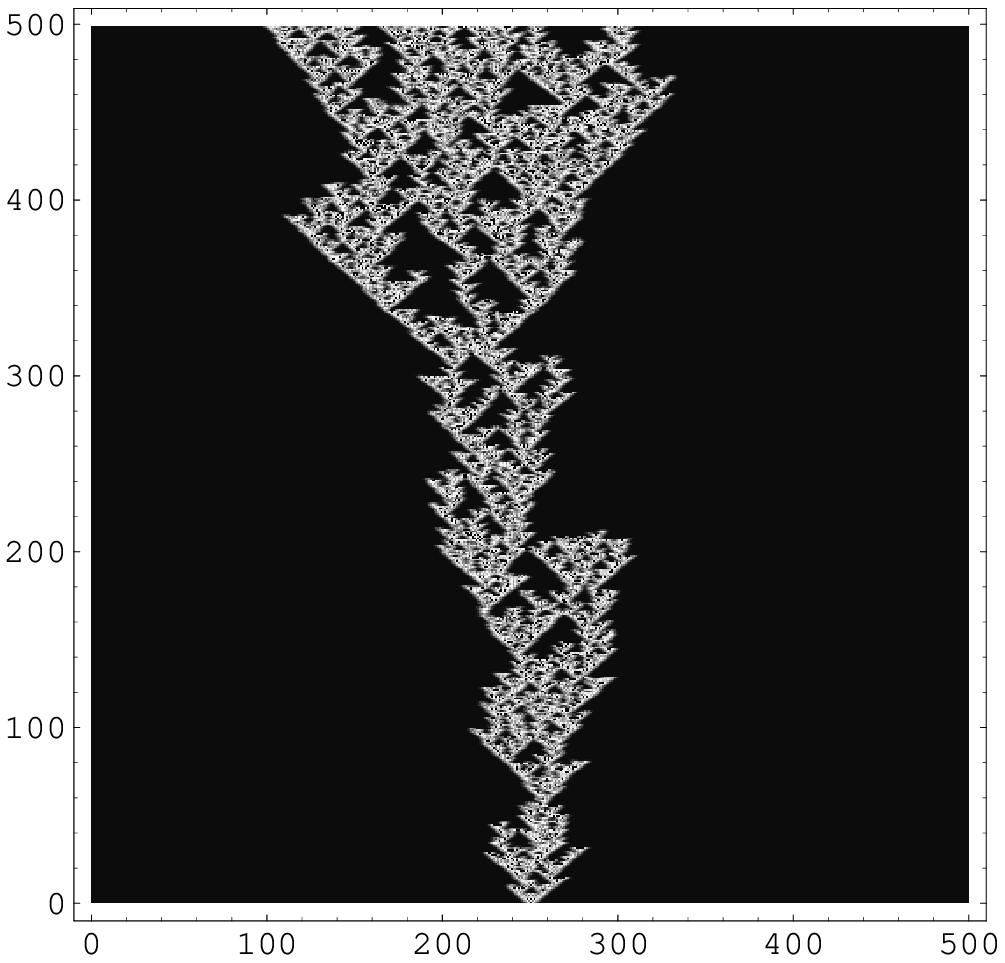}
\caption{We show the spreading of turbulent (white) sites in a laminar (black) background with 2 initial seeds. The horizontal axis is space and time is along the vertical axis. $\Omega=0.049,\epsilon=0.8494$. Lattice size, $L=500$. Every $10^{th}$ iterate has been plotted.}
\end{figure}
\end{center}
The computed values for all the exponents above defined at the critical
values 
$\Omega=0.049, \epsilon=0.8494$ and $\Omega=0.073,
\epsilon=0.4664$ are listed in Table I. It is clear that the calculated
exponents agree very well with the exponents for directed percolation.
Thus the spatio-temporal intermittency at the points indicated by
crosses in the phase diagram (Fig.1) belongs convincingly to the 
directed percolation class.
\begin{table}[!h]
\caption{This table compares static CML exponents with the static DP exponents}
\begin{center}
\begin{tabular}{|c|c|l|l|l|l|l|}
\hline
$\Omega$ &$\epsilon$ & $z$ & $\beta/{\nu z} $ & $\beta$ & $\nu$ & $\eta'$  \\
\hline
0.049 & 0.8494 & 1.578 & 0.16 & 0.278 & 1.1& 1.51 \\
0.073 & 0.4664 & 1.58 &0.16 &0.278 & 1.1 &1.515  \\ 
\hline
{DP}&&1.58&0.16&0.28&1.1&1.51\\ \hline
%\caption{This table compares static CML exponents with the static DP exponents}
\end{tabular}
\end{center}

\end{table}
\begin{table}[!h]
\caption{This table compares dynamic CML exponents with the dynamic DP exponents}
\begin{center}

\begin{tabular}{|c|c|c|c|c|}
\hline
$\Omega$ &$\epsilon$& $\eta$ & $\delta$ & $z_s$\\
\hline

0.049 & 0.8494 &0.313&0.16&1.26\\
0.073 & 0.4664 & 0.313 &0.16&1.26\\ 
\hline
DP&&0.313&0.16&1.26\\ \hline
\end{tabular}
\end{center}
\end{table}

\section{Discussion}

It is interesting to note that the parameter values which correspond to
spatio-temporal intermittency all lie close to the bifurcation
boundaries where bifurcations from synchronised solutions take place.
The uppermost point where spatio-temporal intermittency is seen
corresponds to a tangent-period doubling bifurcation, where two
eigen-values of the stability matric cross $+1$ and $-1$ respectively,
whereas a tangent-tangent bifurcation where several of the eigen values
cross the unit circle takes place at the other values. The synchronised
solutions go to spatio-temporally chaotic solutions via 
spatio-temporal intermittency at these values, whereas they go to spatio-temporal chaos via
cluster solutions in other places. Spatio-tempral intermittency can also
be seen at other locations in the phase diagram. The exact nature of
this behaviour is presently under investigation. We also hope to look at these and 
other aspects of dynamical behaviour in further detail in future work.

\end{document}